\documentclass[twocolumn]{aastex6}
\usepackage{natbib}
\usepackage{enumitem}

\fullcollaborationName{The co-authors}

\shorttitle{Structural Transition in the NGC 6251 Jet}
\shortauthors{Tseng et al.}

\defcitealias{asa12}{AN12}
\defcitealias{nak13}{NA13}

\begin{document}

\title{Structural Transition in the NGC 6251 Jet: An Interplay with the Supermassive Black Hole and Its Host Galaxy}

\author{Chih-Yin Tseng\altaffilmark{1,2}, Keiichi Asada\altaffilmark{1}, Masanori Nakamura\altaffilmark{1}, \\ Hung-Yi Pu\altaffilmark{1}, Juan-Carlos Algaba\altaffilmark{1,3}, and Wen-Ping Lo\altaffilmark{1,2}}

\altaffiltext{1}{Institute of Astronomy \& Astrophysics, Academia Sinica, P.O. Box 23-141, Taipei 10617, Taiwan; cytseng@asiaa.sinica.edu.tw}
\altaffiltext{2}{Department of Physics, National Taiwan University, Taipei 10617, Taiwan}
\altaffiltext{3}{Korea Astronomy and Space Science Institute, 776, Daedeokdae-ro, Yuseong-gu, Daejeon, 305-348, Korea}

\begin{abstract}
The structure of the NGC 6251 jet on the milliarcsecond scale is investigated using images taken with the European VLBI Network and the Very Long Baseline Array. 
We detect a structural transition of the jet from a parabolic to a conical shape at a distance of $(1-2) \times 10^5 $ times the Schwarzschild radius from the central engine, which is close to the sphere of gravitational influence of the supermassive black hole (SMBH). 
We also examine the jet pressure profiles with the synchrotron minimum energy assumption to discuss the physical origin of the structural transition. 
The NGC 6251 jet, together with the M87 jet, suggests a fundamental process of the structural transition in the jets of active galactic nuclei (AGNs). 
Collimated AGN jets are characterized by their external galactic medium, showing that AGN jets interplay with the SMBH and its host galaxy. 
\end{abstract}

\keywords{galaxies: active - galaxies: individual: (NGC 6251) - galaxies: jets - radio continuum: galaxies}

\section{Introduction} \label{sec-intro}
The collimation mechanism of jets from active galactic nuclei (AGNs) remains one of the open questions in high-energy astrophysics. 
Structural studies of the jet in the collimation zone, which requires imaging with (sub-)milliarcsecond resolution, have been uniquely provided by very long baseline interferometry (VLBI). 
It has been found that some of the jets in the nearby sources are gradually collimated on scales of parsecs (e.g., \citealp[M87:][hereafter AN12]{asa12};  \citealp[3C\,84:][]{nag14};  \citealp[Cyg\,A:][]{2016A&A...585A..33B}), while the majority of distant blazar jets exhibit a (freely expanding) conical shape on a scale beyond $\sim$ 10 pc \citep[e.g.,][]{2005AJ....130.1418J, 2014evn..confE.104P}. 
Magnetohydrodynamic (MHD) processes are often invoked to explain their collimation together with acceleration properties \citep{mei01}. 
In principle, the bulk acceleration of MHD jets takes place in a parabolic stream \citep{2006MNRAS.367..375B, 2007MNRAS.380...51K, lyu09}. 
Nevertheless, it is unclear whether the external confinement by the environment plays a crucial role, 
or whether self-collimation by the toroidal magnetic field (i.e., hoop stress) alone can be responsible; 
the latter mechanism is most likely for non-relativistic jets \citep[e.g., jets in young stellar object,][]{con15}. 
 
Recent observations of M87 have found that the structural transition of the jet from a parabolic to conical shape takes place at around the Bondi accretion radius, the outermost extent of the gravitational influence upon ambient gas of the supermassive black hole (SMBH), 
suggesting that the AGN jet collimation is subject to thermal confinement by the stratified interstellar medium (ISM) (\citetalias{asa12}; \citealp[hereafter NA13]{nak13}). 
Interestingly, the simultaneity of gradual acceleration and collimation is also detected in the M87 jet with VLBI monitoring observations (\citetalias{nak13}; \citealp{asa14}). 
However, such a structural study of a jet over multiple orders of magnitude in axial distance has been conducted only for M87 so far. 

NGC 6251 is a nearby \citep[$z=$ 0.025,][]{weg03} giant elliptical galaxy, which has an exceptionally straight and long jet \citep[3 Mpc in projection,][]{wag77}, and therefore serves as one of the best targets for structural studies of AGN jets. 
Its kiloparsec-scale radio emission has been investigated in great detail using the Very Large Array (VLA) \citep{per84}, and the parsec-scale jet emission using VLBI continuously extending to 50\,mas \citep{jon86, sud00, jon02}. 
The mass of the SMBH is measured to be $M = (6\pm2) \times 10^{8}\,M_\odot$ \citep{fer99}. 
Taking the distance ($\simeq$ 103 Mpc) into account, NGC 6251 appears sufficiently large (0.50\,pc\,mas$^{-1}$ or 8700\,$r_s$\,mas$^{-1}$, where $r_s$ is the Schwarzschild radius) for examination of the jet structure across the gravitational regimes dominated by both the SMBH and the host galaxy. 
In this paper, we report a detection of the structural transition of the NGC 6251 jet that is found with the aid of our new images from the European VLBI Network (EVN) at 1.6 GHz. 

The paper is organized as follows. 
In Section \ref{sec-data}, we describe our observations and the data used in this work. 
We show our analysis and results in Section \ref{sec-result}. 
In Section \ref{sec-discuss}, we discuss the location of the transition, the properties of the pressure, and the comparison with M87 in order to understand the origin of the collimation processes. 
The angular scale used in this paper is obtained with a cosmology of $H_{0} =$ 69.6\,km\,s$^{-1}$\,Mpc$^{-1}$, $\Omega_{m} =$ 0.286, and $\Omega_{\Lambda} =$ 0.714 \citep{2006PASP..118.1711W}.

\section{Observations and Data Reduction} \label{sec-data}
  \subsection{EVN Data}
We conducted EVN observations of NGC 6251 on 2013 March 10 at 1.6 GHz with the stations at Badary, Svetloe, Zelenchukskaya (Russia), Effelsberg (Germany), Jodrell Bank (UK), Medicina, Noto (Italy), Onsala (Sweden), Shanghai, Urumqi (China), Torun (Poland), and Westerbork (Netherlands).
Data with both left, and right circular polarization were recorded at each station using eight sub-bands of 8 MHz bandwidth and 2-bit sampling. 
The data were correlated at the JIVE (Joint Institute for VLBI in Europe) correlator. 

Data were calibrated following the standard procedures in AIPS.
A priori amplitude calibration for each station was derived from the system temperatures measured during each run and the antenna gain curves. 
Fringe fitting was performed to remove the residual delays and rates by assuming a point source model.
After applying the solutions, the data were exported to Difmap and averaged over 12 s in each sub-band. 
Then the gain phase and amplitude were self-calibrated by the iterative process of CLEAN to obtain the final image.
More detailed observational parameters are described in Table \ref{tab-1}.

  \subsection{VLBA Data}
Archival Very Long Baseline Array (VLBA) data at 5 GHz are used and calibrated in the same manner as the EVN data.
Also, 12 epochs of the VLBA data at 15 GHz are obtained from the MOJAVE database \citep{lis09}.
Observations were conducted during 1998$-$2013. 
We reproduce all images in agreement with the MOJAVE database.
In order to perform a joint analysis, images of the 12 epochs are convolved with the same circular beam with FWHM of 0.54\,mas (see more details in Table \ref{tab-1}). 

  \subsection{VLA Data} \label{subsec-VLA}
We also use a published VLA image of NGC 6251 at 1.4 GHz to compare with the VLBI measurements in Section \ref{subsec-support}.
The image, as well as the calibration processes, is shown in \citet{sam04}. 
Observations were conducted on 1995 August 15 using the full VLA in its A-configuration.
The beam is restored to be circular with an FWHM of 2\arcsec.

\section{Analysis and Results} \label{sec-result}
  \subsection{New EVN Images}
\begin{figure*}[htb!]
\figurenum{1}
\gridline{\includegraphics[trim={0.3cm 4.3cm 1.2cm 2.5cm},clip=true,width=0.32\textwidth]{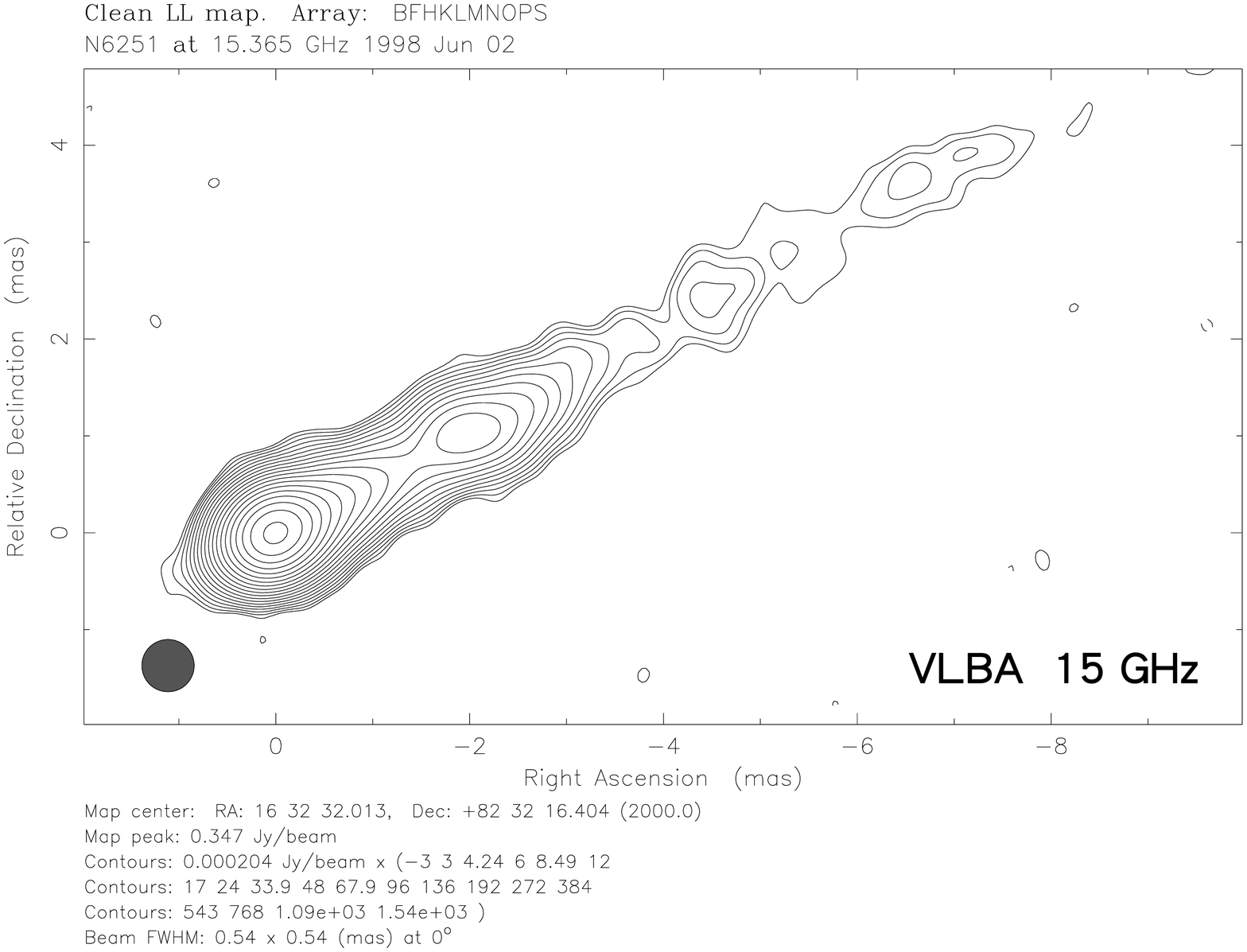}{(a)} }
\gridline{\includegraphics[trim={0.8cm 4.3cm 1.2cm 2.5cm},clip=true,width=0.32\textwidth]{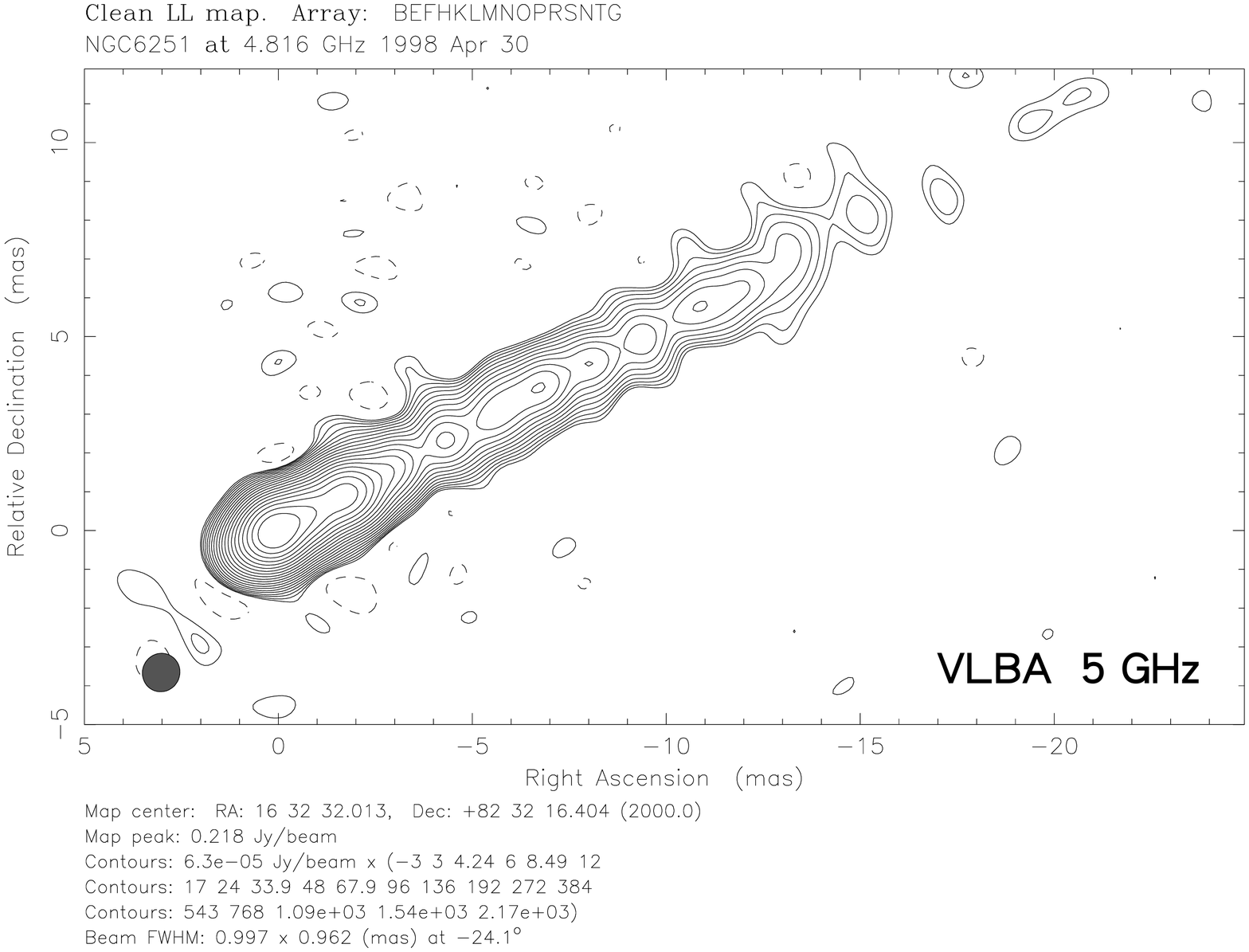}{(b)} }
\gridline{\includegraphics[trim={0.8cm 4.3cm 1.2cm 2.5cm},clip=true,width=0.32\textwidth]{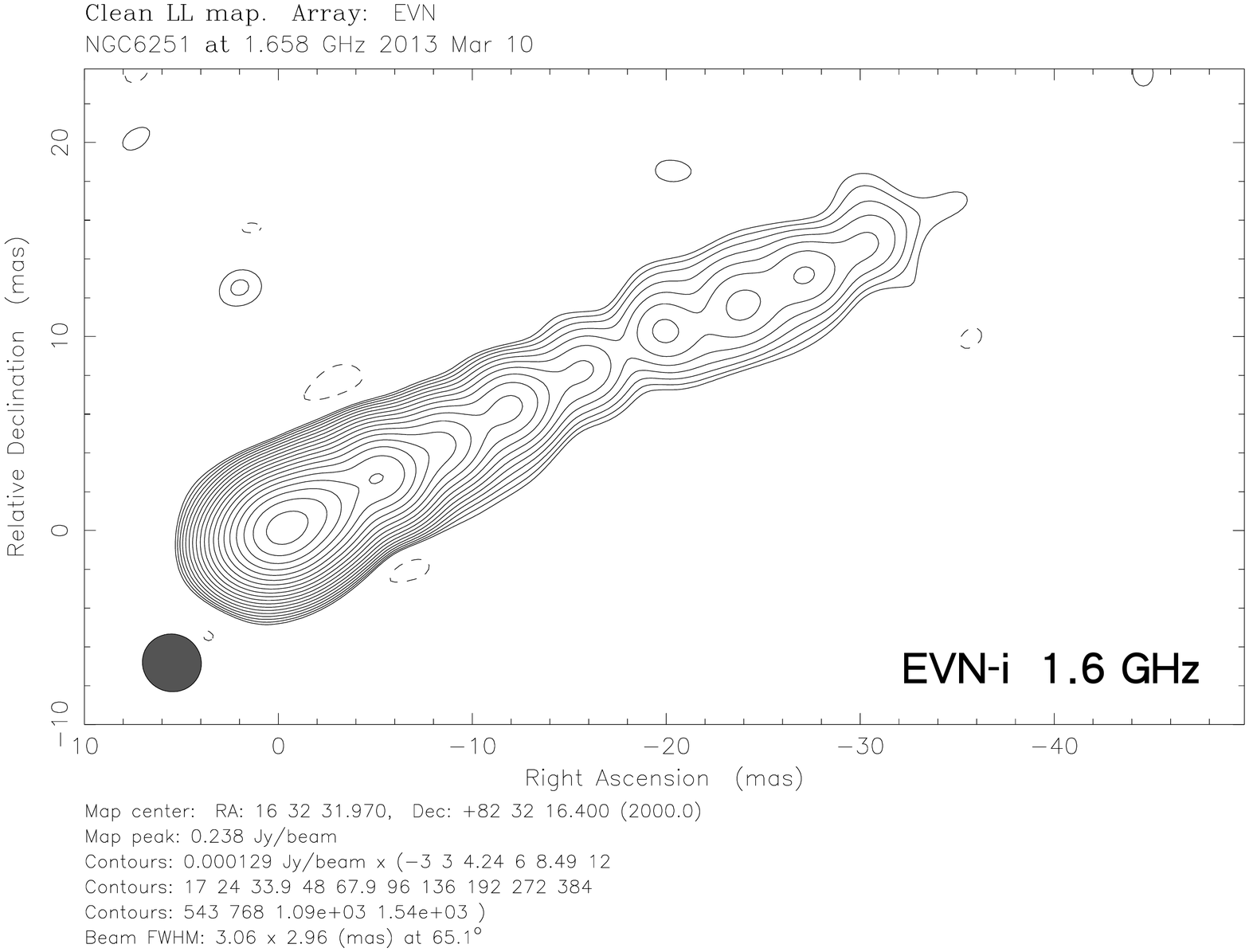}{(c)} }
\gridline{\includegraphics[trim={0.8cm 4.3cm 1.2cm 2.5cm},clip=true,width=0.32\textwidth]{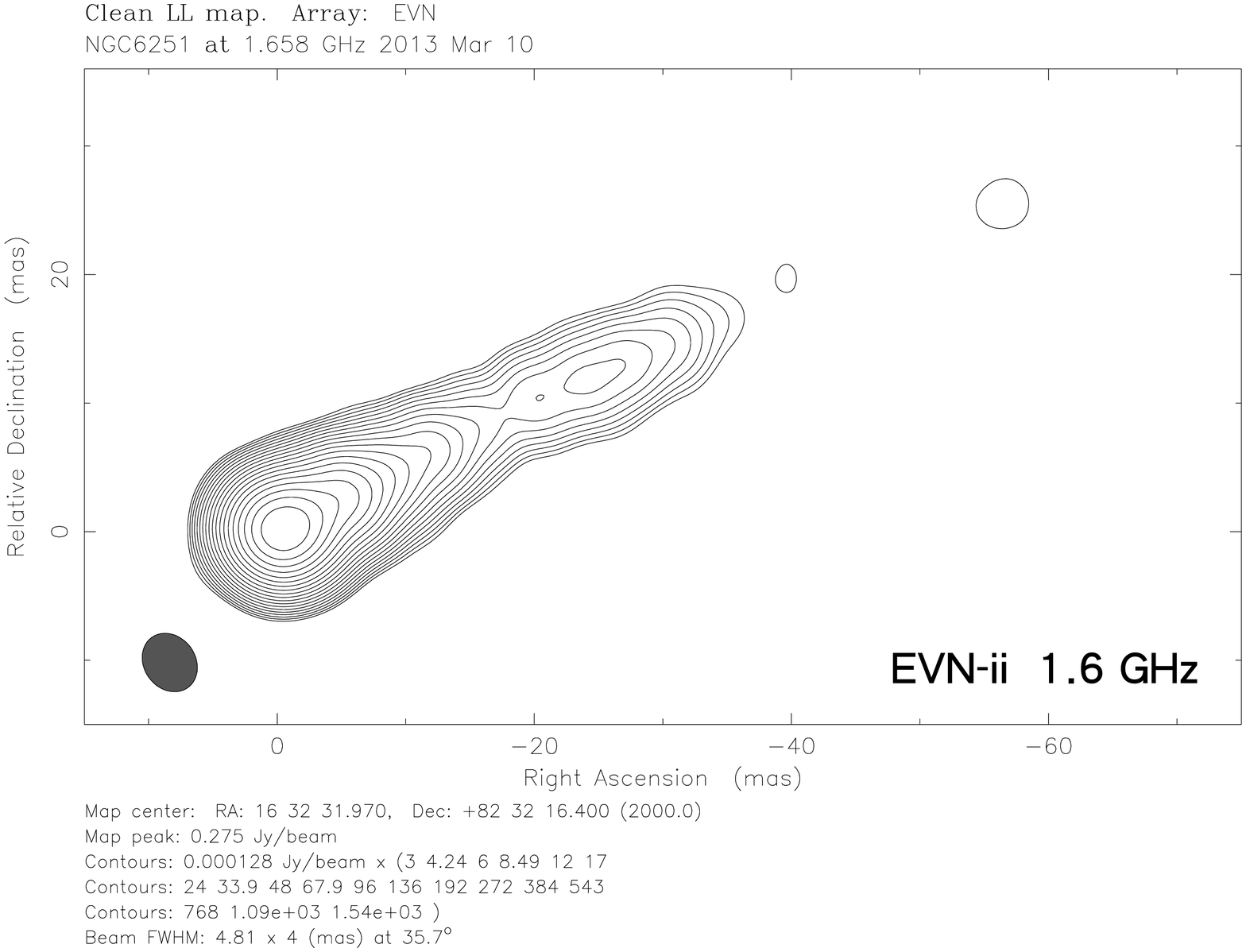}{(d)} }
\gridline{\includegraphics[trim={0.8cm 4.3cm 1.2cm 2.5cm},clip=true,width=0.32\textwidth]{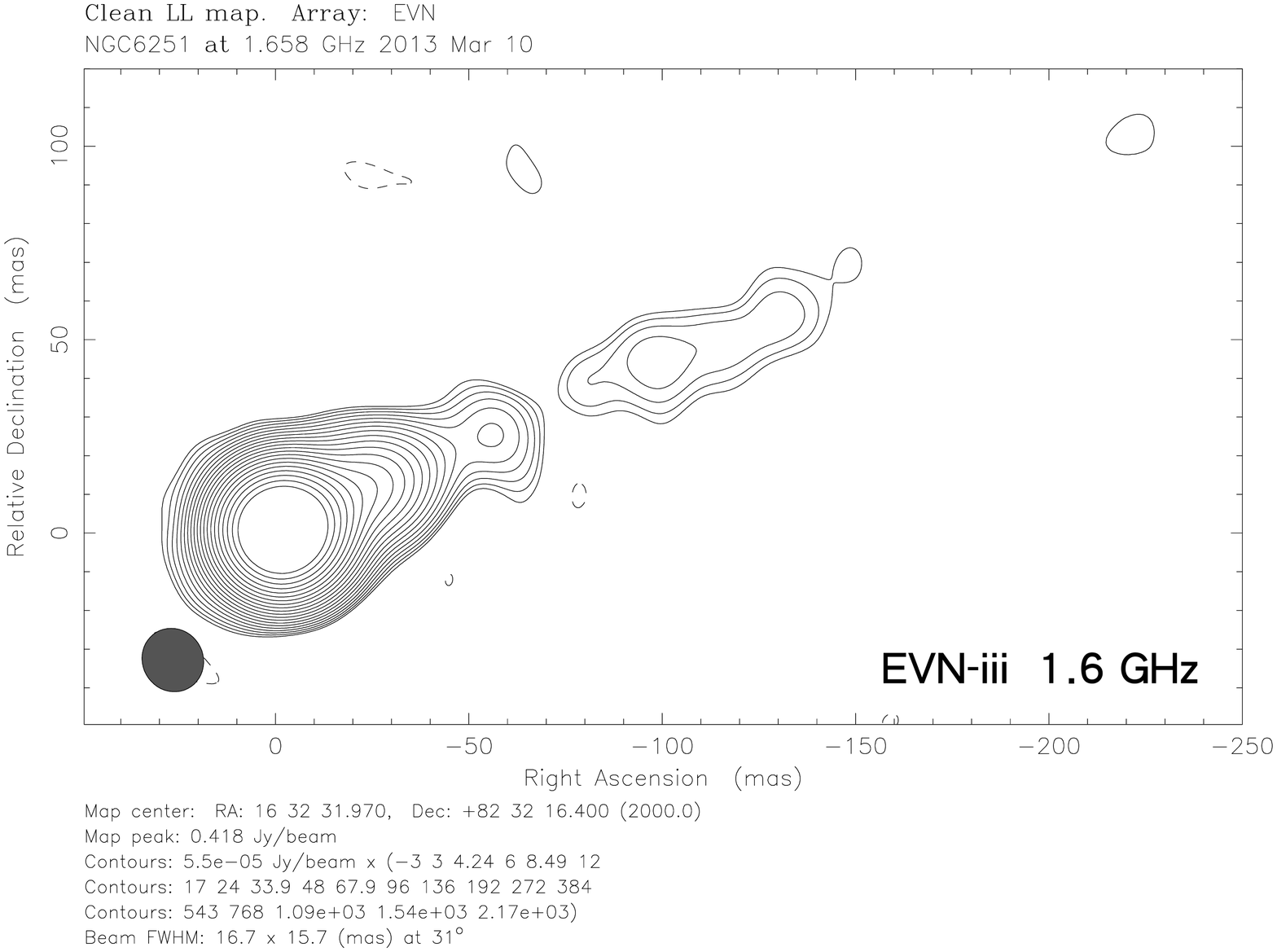}{(e)} }
\caption{
Radio images of the NGC 6251 jet on different length scales. 
From top to bottom, they are taken with 
(a) VLBA at 15 GHz (1998 Jun 02), 
(b) VLBA at 5 GHz, 
(c) EVN at 1.6 GHz with uniform weight (EVN-i), 
(d) EVN at 1.6 GHz with natural weight (EVN-ii), 
(e) EVN at 1.6 GHz with natural weight and tapering (EVN-iii). 
The synthesized beam is shown at the bottom left corner of each image. 
Contours are plotted at ($-$1, 1, $\sqrt{2}$, 2, ...) $\times$ 3$\sigma$, where $\sigma$ is the residual rms noise. 
See the values of $\sigma$ and the beam sizes in Table \ref{tab-1}. 
Note that the intrinsic jet radius (or width) is obtained by deconvolution from the beam, which is not directly illustrated from the contour level. 
  \label{fig-image} }
\end{figure*}

Figure \ref{fig-image} shows images of NGC 6251 with high dynamic range taken from our EVN observations at 1.6 GHz, and the archival VLBA data at 5 and 15 GHz; Table \ref{tab-1} summarizes the detailed image parameters (e.g., beam size, dynamic range). 
The EVN images are obtained with different weighting schemes (denoted as EVN-i, -ii, and -iii) to enhance the sensitivity of the emission on various angular scales. 
The brightest component at the eastern edge of the object is presumably the core, and a continuous jet emission extends towards the northwest. 
The jet direction and the knotty features are consistent with each other in all images. 
This is the first time that the jet emission associated with NGC 6251 has been continuously detected at a distance range of 50$-$150\,mas away from the core. 

  \subsection{Derivation of Jet Radius}
To investigate the collimation profile of the jet, we derive the radius (half width) of the jet emission along the jet axis in each image. 
We mainly adopt the methodology of AN12 as follows; in addition, the change in the jet direction on different length scales (i.e., jet bending) is taken into consideration: 
\begin{enumerate}

\item 
The position angle (PA) of the jet direction is determined from the averaged PAs of all CLEAN components excluding the core region, weighted by their flux density. 

\item
The cross sections of the jet are fitted with a single Gaussian function of FWHM $\Phi_0$ at every 1/5 of the beam size (as independent measurements, the resolution limit was studied by \citealt{2005astro.ph..3225L}) along the jet axis to evaluate the transverse structure. 

\item
The jet radius $r$ (i.e., half of the jet width) is thus defined as half the FWHM of the resultant fitted Gaussian, which is deconvolved from the beam FWHM $\Phi_b$ taking into account the beam orientation, i.e., $2r = \sqrt{\Phi_0^2 - \Phi_b^2}$ \citep[see also][Section IV-d]{per84}. 

\item 
The uncertainty is estimated from the Gaussian fitting error (in step 2), and the imaging error that is evaluated as: 
  \begin{enumerate}

  \item
  The dispersion (standard deviation) of the 12-epoch measurements for the VLBA data at 15 GHz.

  \item
  $\Phi_b$/5 for the other frequencies due to the lack of sufficient independent observations. 
  Particularly for the EVN data, the jet radii are further averaged over both polarizations (i.e., RR and LL).

  \end{enumerate}

\end{enumerate}
The measurements are presented in a machine-readable format in Table \ref{tab-2}. 
In our analysis, the jet PA decreases mildly as the length scale increases, ranging from 294$\degr-$298$\degr$ (see also Table \ref{tab-1}); a similar decrease is detected by \citet{jon02}.
All jet cross sections are well represented with a single Gaussian component. 

To properly evaluate the uncertainty, we utilized the multi-epoch VLBA data at 15 GHz. 
The jet radii derived in 12 epochs are in good agreement with each other so that we obtain the standard deviation ($\sigma_{r,15}$) of the jet radii as the imaging error. 
We estimated the distribution of $\sigma_{r,15}$ as a function of signal-to-noise ratio (S/N), and found it to be well described by a power-law function, $\sigma_{r,15}/\Phi_b =$ (0.66 $\pm$ 0.09) $\times$ S/N$^{-0.68\pm0.05}$. 
The $\sigma_{r,15}$ is about 0.01\,mas at the innermost jet region where the S/N $\simeq$ 200, and 0.1\,mas downstream of the jet where S/N $\simeq$ 5, corresponding to 1/50 and 1/5 of the beam size, respectively. 
Therefore, we consider the upper limit of the imaging error to be $\Phi_b$/5 because the jet radii are measured where the jet is detected when S/R $\geq 5$ \citep[see also][Appendix A for similar analysis]{hom02}. 
Note that we assume the time variability of the jet radius is negligible. 

  \subsection{Radius Profile of the Jet}
Figure \ref{fig-rz} shows the distribution of jet radius ($r$) as a function of the deprojected distance to the core ($z$), which we refer to as the radius profile of the jet. 
The jet viewing angle ($\theta$) for deprojection is uncertain within an upper limit of $47\degr$ \citep{jon02}. 
We adopt $\theta =$ 19$\degr$ derived from a detection of a sub-parsec scale counter-jet \citep{sud00}. 
An independent model of the spectral energy distribution (SED) suggests a consistent result of $\theta \simeq 18\degr$ \citep{chi03}. 
We use the deprojected length scale ($27000\,r_s$\,mas$^{-1}$) along the jet throughout the rest of the paper. 
Note that the jet radius, as defined perpendicular to the jet axis, is not affected by projection effects.

  \subsection{Collimation Profile by Power-Law Models}
Collimation profiles are generally described by power-law functions with some local features, as a manifestation of the global evolution of jet structure. 
We perform regression analyses with single and double power-law models throughout the whole VLBI data set and show the results and their relative residuals in Figure \ref{fig-rz}. 
Note that the power-law indices, $a$, are defined as \ $r \propto z^{1/a}$ \ throughout the paper. 

First, we fit the data with a single power-law model, $r_{SP}$, 
  \begin{equation}
r_{SP}(z) = A_0\,z^{1/a},
  \end{equation}
which gives the best fit using a power-law index of $a =$ 1.25 $\pm$ 0.03, with a $\chi^2$/dof $\simeq$ 0.98 (see also Table \ref{tab-3}). 
The radius profile can be described well with this power law in the range of $z = (10^5 - 10^6)\,r_s$, but it deviates systematically at each end. 

Then, we test a double power-law model, in the form of the broken power-law function, $r_{DP}$, 
  \begin{equation}
r_{DP}(z) = r_0 \left[ \left(\frac{z}{z_0}\right)^{n/a_u} + \left(\frac{z}{z_0}\right)^{n/a_d} \right] ^{1/n},
  \end{equation}
so that we can simultaneously solve for $a_u$, $a_d$, $z_0$ and $r_0$, which are the upstream and downstream power-law indices (pre- and post-break), the break position, and the jet radius at the break position, respectively, and $n$ is a controlling parameter for the sharpness of the break, i.e., the sharper the break, the larger $n$. 
The radius profile is best fitted with the model parameters listed in Table \ref{tab-3}, with $\chi^2$/dof $\simeq$ 0.57$-$0.58 by adopting different $n\geq1$. 

Note that the different values of sharpness (i.e., choices between $n$) result in changes of the best fitted parameters (see also Table \ref{tab-3}). 
In particular, a small value of $n \gtrsim 1$ implies a smooth transition with a significant amount of data located amid the two pure power-law branches, causing an increase in the uncertainty for both $a_u$ and $z_0$. 
However, the resultant fittings are all fairly consistent (in 1\textit{$\sigma$}, and in 2\textit{$\sigma$} for the cases $n=1$ and $n=2$) with power-law indices $a_u \geq$ 2 and $a_d =$ 0.9$-$1, and the distance of the transition (break position) $z_0 =$ (1$-$2) $\times$ 10$^5\,r_s$ regardless of the sharpness. 
In addition, their relative residuals are uniformly distributed on all distance scales. 
The oscillatory pattern in the radius profile, presumably corresponding to a local enhancement of the emissivity, potentially affects the quantification of the power-law index of the upstream. 
In order to have a better constraint, it is important to extend the structural studies of the jet towards the inner regions. 
Future (sub)millimeter VLBI observations, with unprecedented angular resolution, are expected to provide a unique chance to unveil the initial point of collimation and even the genesis of AGN jets. 

Next, we perform a statistical test (an $f$-test) to check whether the double power-law model is necessary. 
Given the $f$-ratio of 1.71 (i.e., the ratio of $\chi^2$/dof) with the degrees of freedom of the single and double power-law fitting, $dof=$ 189 and 187, respectively, 
the null hypothesis that the two models fit the population of the data equally well corresponds to a small probability $p \sim 1.4 \times 10^{-4}$. 
As a result, the double power-law model is significantly better than the single power law at describing the radius profile of the NGC 6251 jet. 
Also, this is further supported by the measurements in both the outer and inner regions taken with the VLA and VLBI core data (see Section \ref{subsec-support}). 

In summary, we conclude that the structure of the NGC 6251 jet is described by a parabolic shape ($a\simeq$ 2) upstream, a conical expansion ($a\simeq$ 1) downstream, and the structural transition takes place at a characteristic distance scale of $(1-2)\times 10^5\,r_s$. 
The NGC 6251 jet is the second piece of observational evidence of the structural transition from parabolic to conical shape occurring at a scale of 10$^5$ $r_s$ in AGN jet systems, following the result of M87 \citepalias{asa12}.

\section{Discussion} \label{sec-discuss}
\subsection{Evidence Supporting the Structural Transition} \label{subsec-support}
\begin{figure*}[htb!]
\figurenum{2}
\includegraphics[angle=270, width=0.6\textwidth]{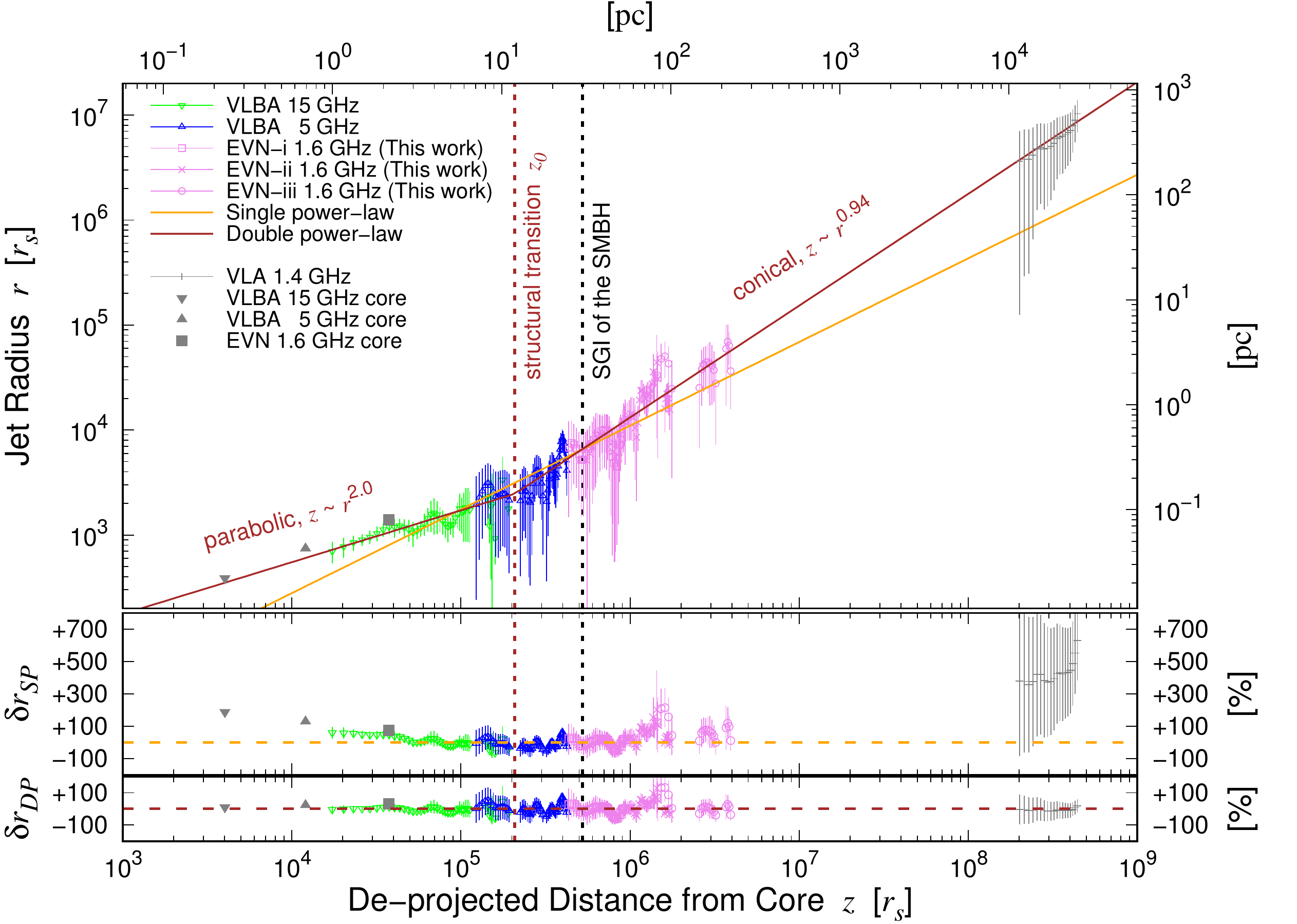}
\caption{
Top panel:  the radius profile of the NGC 6251 jet in units of Schwarzschild radius ($r_s$) and parsecs, using archival VLBA data at 15 and 5 GHz (green and blue triangles), and our EVN(-i, ii, iii) data at 1.6 GHz with three weighting schemes (violet square, cross, and circle). 
The single (orange solid line) and double (brown solid line) power-law models are fitted to the profile, where we show the case of the sharpness parameter $n=$ 100 here. 
Statistics ($f$-test) shows that the data are better fitted by double power-law models (with all $n$) than by a single power law (see also Section \ref{sec-result}). 
Furthermore, the double power-law model is also supported by the VLA data (gray plus signs) and VLBI cores (gray triangles and squares), as discussed in Section \ref{subsec-support}. 
Note that the VLA data and VLBI cores are not included in the regression analysis. \\
Middle and bottom panels:  the relative residuals of the jet radii for the single ($\delta r_{SP} = r/r_{SP}-$ 1) and double ($\delta r_{DP} = r/r_{DP}-$ 1; $n=$ 100) power-law models. 
The single power law shows a systematic deviation at the two ends of the distance scale, while the double power laws (i.e., for all $n$) show uniformly distributed relative residuals on all distance scales. \\
We conclude that the jet structure is described by a parabolic shape upstream ($z \propto r^{a}, a \simeq 2$) and a conical shape downstream ($a \simeq 1$) with a transition located at a characteristic distance $z_0=$ (1$-$2) $\times$ 10$^5$ $r_s$ (brown dashed line), 
which is fairly close to the sphere of gravitational influence (SGI) of the SMBH ($r_{SGI} \simeq$ 5 $\times$ 10$^5$ $r_s$, black dashed line) in order of magnitude. 
  \label{fig-rz} }
\end{figure*}

We take a step further to investigate the jet structure beyond the two ends of the distance scale, which are shown as the gray data points in Figure \ref{fig-rz}. 
The radius profile on kiloparsec scale is derived from an archival VLA image (as introduced in Section \ref{subsec-VLA}), which was similarly measured by \citet[Fig. 10]{per84}. 
Figure \ref{fig-rz} shows their innermost portion (with distances $z \leq$ 20 kpc) to guide the comparison; this portion is smoothly connected to the radius profile predicted by the double power-law model, while largely deviating from the single power-law model. 
The VLA measurements additionally support the idea that the VLBI jet downstream is better described by a conical shape ($a_d\sim 1$) that propagates three orders of magnitude in distance, if the jet structure is indeed linked together on the intermediate scales ($z=$ 10$^6-$10$^8$ $r_s$). 
Note that the VLA data are not included in the previous regression analysis. 
To study the missing link of the jet structure requires facilities with an adequate angular resolution (e.g., VLA, MERLIN), which will be investigated in future work. 

On the other hand, VLBI cores are considered to be the innermost jet emissions, as originally suggested by \citet{BK79}. 
The frequency-dependent core-shift ($\Delta_z \propto \nu^{-1}$) due to synchrotron self-absorption enables us to estimate the exact location of the innermost components with respect to the black hole, and hence the innermost jet structure and its width.
This technique has been previously demonstrated in M87 (e.g., \citealp{had11, 2013ApJ...775...70H}, \citetalias{nak13}). 
In NGC 6251, the core-shift between 15 and 5 GHz was estimated to be $\Delta_z \simeq$ 0.3\,mas by image registration with the optically thin jet \citep{sud00}. 
Together with the transverse widths (FWHM, deconvolved from the beam) of the cores, the sub-parsec-scale jet radii can be estimated by the VLBA and EVN cores (see also Figure \ref{fig-rz}). 
Once again, they are relatively consistent with the upstream region that is predicted by the double power-law model rather than the single power law. 
As a result, we expect that the jet maintains a parabolic shape over at least two orders of magnitude in distance. 
Interestingly, the power-law index of the jet upstream ($a_u\sim2$) is very close to the genuinely parabolic field configuration in which the jet extracts electromagnetic energy from a spinning black hole \citep{BZ77}. 
Core-shift measurements of NGC 6251 at higher frequencies (i.e., submillimeter bands) can be very helpful in understanding the jet formation process as well as in extending the structural study. 

\subsection{Location of the Structural Transition}
We consider the gravitational potential of the environment of NGC 6251: the SMBH and the host galaxy. 
As indicated in Figure \ref{fig-rz}, it is found that the structural transition is close to the boundary of the sphere of gravitational influence (SGI, $r_{SGI} \simeq$ 5 $\times$ 10$^5$ $r_s$) in order of magnitude, where the central SMBH dominates the gravitational potential. 
The SGI is defined by $r_{SGI}=GM/\sigma_v^2$ under the assumption that the stellar distribution is supported by random motion rather than rotationally, where $\sigma_{v}$ is the velocity dispersion of the collective stars, i.e., the elliptical galaxy here \citep{1972ApJ...178..371P}. 
It is measured by optical spectroscopy that $\sigma_v \simeq$ 290 $\pm$ 14 km\,s$^{-1}$ in NGC 6251, and the ratio of rotational velocity to velocity dispersion is $v/\sigma_{v} =$ 0.16 \citep{fer99, 2013ARA&A..51..511K}, which gives our estimates of the $r_{SGI}$ above. 
The estimated SGI can be an upper limit, since one may expect much larger $\sigma_v$ towards the nucleus of NGC 6251, for we cannot spatially resolve the region at present. 
Note that the evaluation of the SGI may be affected by the triaxiality of the bulge, or the warped disky structure in the nucleus possibly due to galaxies merging, as discussed in \citet{fer99}. 
A study of the stellar velocity structure with an angular resolution $\sim$10\,mas will be helpful in narrowing down this issue. 
However, we believe that both effects are insignificant in order-of-magnitude estimates. 
Approximately, the cool core is expected to be thermodynamically stable, so that virial equilibrium is considered for the gas as well as for the stars. 
As a consequence, the Bondi radius ($r_B=2GM/c_s^2$, where $c_s$ is the local sound speed) is expected to be close to the SGI. 
For instance, in M87, which is the very limited case that we can directly compare those two numbers with thanks to its proximity, the Bondi radius is indeed close to the SGI ($r_B \simeq$ 3.8 $\times$ 10$^5$ $r_s$, \citetalias{asa12} and the references therein, and $r_{SGI} \simeq$ 3.1 $\times$ 10$^5$ $r_s$ with $\sigma_v \simeq$ 380 km\,s$^{-1}$, \citealp{2014ApJ...785..143M}; we adopt an SMBH mass of $6.6 \times 10^{9}\,M_\odot$, \citealp{2011ApJ...729..119G}). 

Therefore, the Bondi radius in NGC 6251 is expected to be a few times 10$^5\,r_s$ with a virial temperature $k_BT_{vir} \simeq \mu m_p\sigma_v^2\simeq$ 0.5 keV derived from the velocity dispersion, where $\mu\ (\simeq 0.6)$ and $m_p$ are the mean molecular weight and proton mass. 
Independently, we estimate $c_s = \sqrt{\gamma p/\rho} = \sqrt{\gamma k_BT/m_p} \simeq$ 500 km\,s$^{-1}$ with an upper limit of temperature $kT \simeq$ 1.6 keV, which is extrapolated from X-ray observation on a scale of 20 kpc \citep[or 10\arcsec,][]{sam04}, where $\gamma$ (= 5/3), $p$, and $\rho$ are the adiabatic index, pressure, and density of the gas. 
This estimation gives the lower limit of the Bondi radius as $r_B \simeq$ 4 $\times$ 10$^5$ $r_s$ and agrees well with the former expectation. 
Note that the extrapolation is made towards the Bondi radius assuming a flat temperature profile; it was demonstrated in many simulations that the profile is nearly flat in the cool core of elliptical galaxies \citep[e.g.,][]{2012MNRAS.424..190G, 2013MNRAS.432.3401G}. 

\begin{figure*}[htb!]
\figurenum{3}
\includegraphics[angle=270, width=0.6\textwidth]{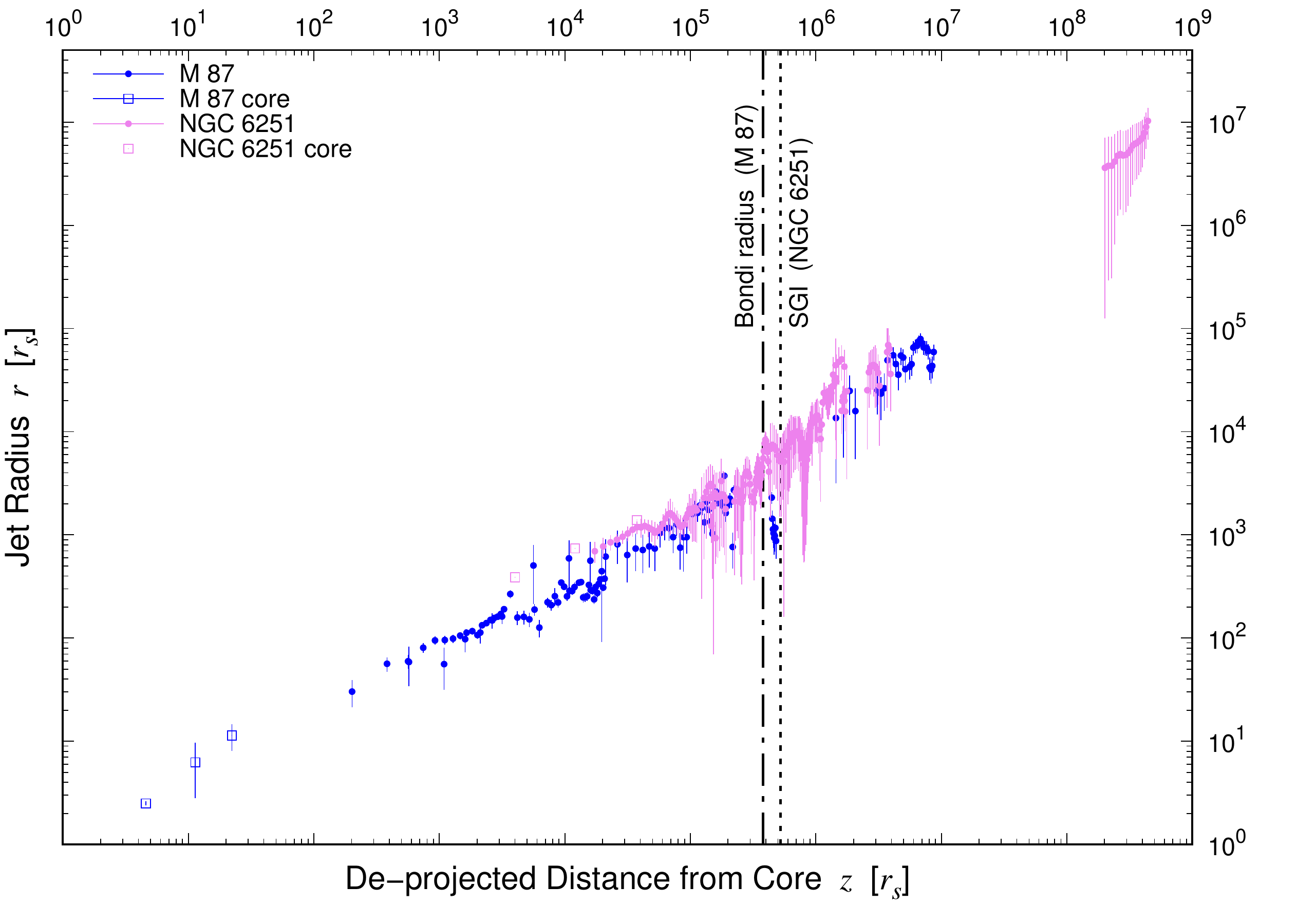}
\caption{
Combined radius profiles of the jets in NGC 6251 and M87, with their jet radii and deprojected distances normalized by their Schwarzschild radii ($r_s$). 
The detailed description of the measurements can be found in Figure \ref{fig-rz}, \citetalias{asa12} and \citetalias{nak13}. 
Likewise, M87 shows a parabolic shape ($z \propto r^{a}, a \simeq$ 1.7) in the upstream region, a conical shape ($a \simeq$ 1) in the downstream region, and a structural transition at a few 10$^5$ $r_s$, corresponding to the Bondi radius of M87 $r_B \simeq$ 3.8 $\times$ 10$^5$ $r_s$ (dotted-dashed line), while the SGI of NGC 6251 is $r_{SGI} \simeq$ 5 $\times$ 10$^5$ $r_s$ (dashed line). 
The similarities of the two jet structures suggest that the collimation process of AGN jets is fundamentally characterized by their external galactic medium, which is gravitationally controlled by the SMBH and its host galaxy. 
  \label{fig-comparison} }
\end{figure*}

We compare the collimation (radius) profile of the jet in NGC 6251 to that in M87, because the structural transition of the M87 jet is also located at around the Bondi radius \citepalias{asa12}. 
Figure \ref{fig-comparison} shows their jet radii and deprojected distances normalized respectively by their Schwarzschild radii. 
Located in the same type of galaxy, the two jets show remarkably similar structures. 
They both undergo a structural transition at a scale of $10^5\,r_s$, suggesting an interplay with the galactic environment, i.e., the SMBH and its host galaxy. 
The collimation process of AGN jets may be fundamentally characterized by the external galactic medium, for which it has been argued that the stratified ISM is responsible \citepalias{asa12, nak13}. 
Recently, it has also been shown that the jet can be confined by the wind generated from the accretion flow within the Bondi radius \citep{2015ApJ...804..101Y, 2016arXiv160407408G}. 

\subsection{Pressure Estimates from Synchrotron Minimum Energy Assumption} \label{subsec-sme}
The internal pressure of the jet is estimated in order to compare it with the external medium. 
We consider the NGC 6251 jet to be a relativistic system organized by helical magnetic field, 
i.e., the toroidal component of the field (denoted by $B_\phi$ in the observer's frame, and $B{'}_\phi = B_\phi/\Gamma$ in the fluid comoving frame, where $\Gamma$ is the bulk Lorentz factor) is dominant farther from the black hole. 

Assuming the jet satisfies the synchrotron minimum energy condition \citep{pac70}, the magnetic field strength $B_\phi$ is estimated as 
\begin{equation}
B_{eq} = \left[\ 6\pi (1+k)\, c_{12}(\alpha, \nu_{min}, \nu_{max})\, L_{syn}(\alpha, \nu_{min}, \nu_{max}) / (\Phi V)\ \right]^{2/7}, 
\end{equation}
(equivalently, the internal pressure $p_{eq}\simeq B_{eq}^2/8\pi$) by using the radio emissivity along the jet. 
First, we adopt a spectral index $\alpha \simeq$ 0.6 ($S_\nu \propto \nu^{-\alpha}$) that has been found in the jet at both kiloparsec  scale and parsec scale (optically thin) \citep{per84, jon02}, 
and a nominal frequency integrand of 10$-$10$^5$ MHz to estimate the synchrotron luminosity $L_{syn}$, as well as to determine the factor $c_{12}$; 
we set the proton-to-electron energy ratio $k$=1 as a strict minimum; 
the emitting volume $\Phi V$ is estimated with a filling factor $\Phi$=1 and a circular jet cross section characterized by the jet radius, which is derived in Section \ref{sec-result}. 
Second, we estimate the bulk Lorentz factor $\Gamma$ by adopting an empirical correlation $\Gamma\,\theta_{j} \simeq$ 0.2 with the half opening angle $\theta_{j}=\tan^{-1}(r/z)$, which has been found in large samples of AGN jets \citep{2005AJ....130.1418J, 2009A&A...507L..33P, 2013A&A...558A.144C}. 
This relation is also demonstrated in the MHD models, indicating a causally connected jet in the acceleration zone \citep[e.g.,][]{kom09, 2009ApJ...699.1789T, 2015ApJ...801...56P}. 
As a result, Figure \ref{fig-pressure} shows the field strength ($B{'}_{eq}=B_{eq}/\Gamma$) together with the jet pressure ($p'_{eq} = B{'}_{eq}^2$/8$\pi$) in the fluid frame. 
Note that true minimum-energy field strength can be one order of magnitude higher than the strict lower limit by tuning the parameters mentioned above. 

\subsubsection{The Parabolic Jet}
To maintain a parabolic jet ($a\simeq$ 2) in the upstream region, pressure matching is expected between the jet boundary and the external ISM ($p'_{jet} \simeq p_{ext}$). 
Analytical and numerical models have shown that the magnetized jet can be genuinely parabolic when the ISM pressure has a profile of $p_{ext} \propto z^{-b}$, $b=$ 2 \citep{zak08, kom09}, plotted as the black dotted line in Figure \ref{fig-pressure}. 
Interestingly, this is consistent with the estimated minimum-energy jet pressure profile if there is no significant change in the synchrotron parameters. 
In addition, at the structural transition, the estimated jet pressure $p'_{jet} \sim$ of a few $10^{-9}$\,dyn\,cm$^{-2}$ is in agreement with the external ISM pressure at the SGI (or the Bondi radius), $p_{ext} = n_e k_B T\simeq$ 10$^{-10}-$10$^{-8}$ dyn\,cm$^{-2}$, where we adopt a nominal range of electron number density in typical radio galaxies $n_e =$ 0.1$-$10 cm$^{-3}$ \citep{1406.6366}. 
As a result, it is reasonable to believe that the external pressure of the parabolic jet has a profile of $b\simeq$ 2, which is consistent with models of radiative inefficient accretion flow (RIAF) (1.5 $\leq b \leq$ 2.5, \citetalias{nak13} and references therein). 
Similar discussions can be referred to in \citetalias{asa12} and \citetalias{nak13}. 
Note that the accretion type of the SMBH in NGC 6251 remains a matter of debate: RIAF or standard accretion disk. 
\citet{1999ApJ...516..672H} suggested that an advection-dominated accretion flow (ADAF, a subtype of RIAF) is present in the nucleus of NGC 6251 based on the radio-to-X-ray SED. 
On the other hand, the X-ray spectral properties are in favor of the presence of a standard accretion disk \citep{2004A&A...413..139G}. 
Our result supports RIAF as being the accretion type in NGC 6251. 

\subsubsection{The Conical Jet}
Beyond the structural transition, the jet changes its structure to a conical shape ($a\simeq$ 1). 
We consider two scenarios for forming a conical jet regarding the pressure properties at the structural transition, which we call a sharp transition ($p'_{jet} > p_{ext}$) and a smooth transition ($p'_{jet} \simeq p_{ext}$). 

\begin{figure*}[htb!]
\figurenum{4}
\centering
\includegraphics[angle=270, width=0.6\textwidth]{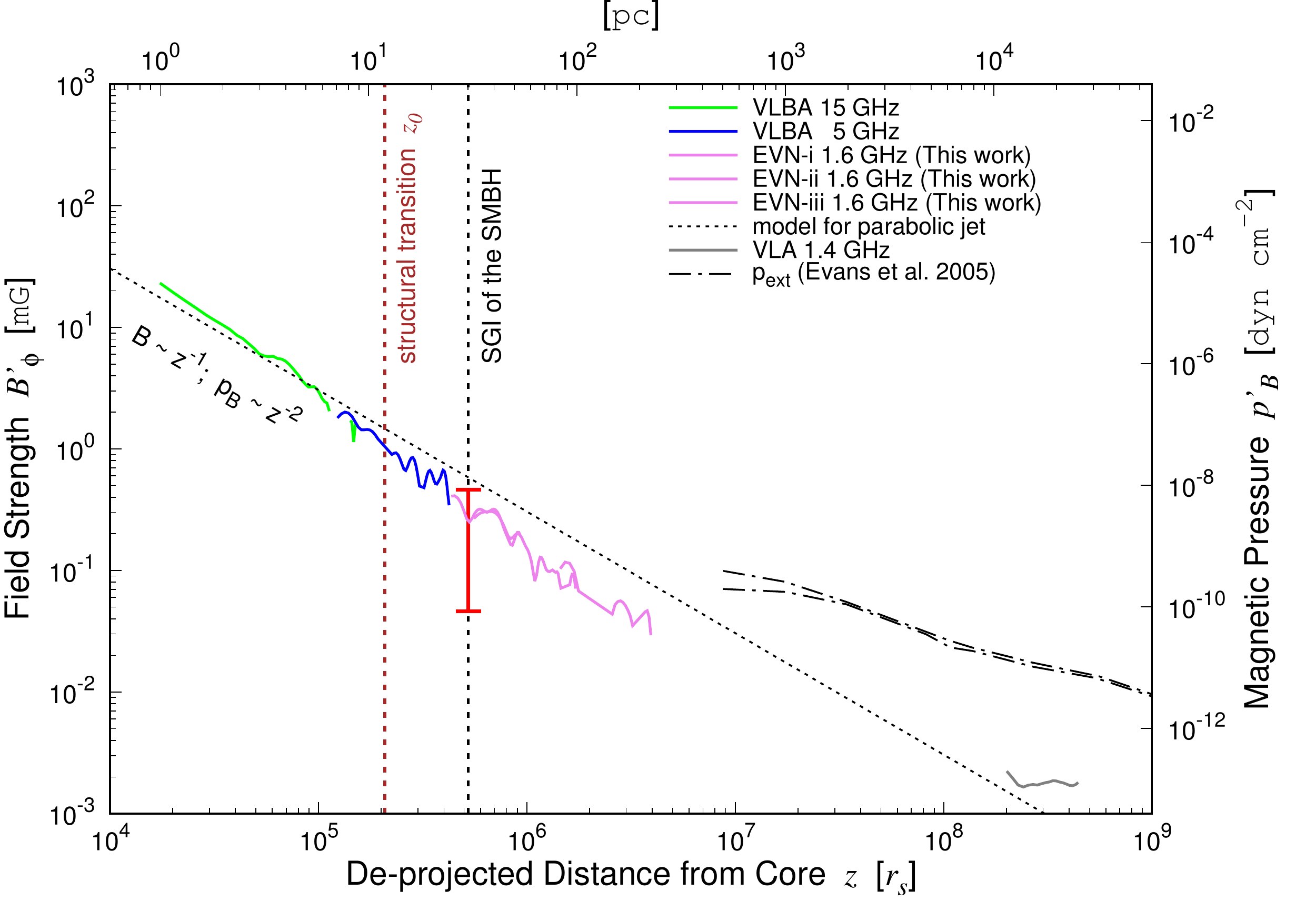}
\caption{
Distribution of the lower limit of the (fluid frame) magnetic field strength $B{'}_{eq}$ and the equivalent magnetic pressure $p{'}_{eq} = B{'}_{eq}^2$/8$\pi$ of the NGC 6251\, jet as a function of the deprojected distance to the core $z$. 
The synchrotron minimum energy condition and an empirical relation $\Gamma\,\theta_{j}=$ 0.2 \citep{2013A&A...558A.144C} are assumed, where $\theta_{j}$ is the intrinsic half opening angle. 
The vertical dashed lines are as defined in Figure \ref{fig-rz}. 
The dotted line illustrates the theoretical profile of the external pressure $p_{ext} \propto z^{-b}, b=$ 2 for a parabolic MHD jet \citep{zak08, kom09}. 
The red line segment indicates the range of the external ISM pressure at the SGI, $p_{ext} = n_e k_B T\simeq$ 10$^{-10} -$ 10$^{-8}$ dyn cm$^{-2}$, with a nominal range of electron number density in typical radio galaxies $n_e =$ 0.1$-$10 cm$^{-3}$ \citep{1406.6366}. 
The two dotted-dashed lines represent the 1$\sigma$ uncertainties of the external gas pressure $p_{ext}$ found by X-ray observation \citep{eva05}. 
See Section \ref{subsec-sme} for more discussions. 
  \label{fig-pressure} }
\end{figure*}

A sharp transition is where the jet becomes overpressured with respect to the external pressure at the transition while it undergoes an overcollimation (due to a compression shock) and adopts a conical shape. 
In the case of M87, the knot complex (HST-1), which is located at the structural transition in the M87 jet, is likely caused by an imbalance between $p'_{jet}$ and $p_{ext}$ (\citetalias{asa12} and references therein). 
As Figure \ref{fig-pressure} shows, $p_{ext}$ is nearly a constant in the conical jet regime, as suggested by the estimation of the SGI and by X-ray measurements \citep{eva05}, which is expected in the galactic environment beyond the SGI. 
It has been shown by \citet{1986ApJ...311L..63C} in an MHD simulation that an overpressured jet can be conically expanding in a flat external medium. 
However, it remains unclear whether the jet is overpressured because the lower limit of the minimum-energy jet pressure is comparable to the external pressure. 
We note that $\Gamma$ (or the jet speed) is one of the important parameters, and it is assumed to follow $\Gamma\,\theta_{j} \simeq$ 0.2, i.e., no deceleration in the conical jet. 
In the case of M87, the jet has also shown to have a peak apparent speed at the transition \citep{asa14}. 
It is interesting that the kiloparsec-scale pressure estimates of NGC 6251 show a significant difference between the estimated jet pressure (given by the VLA measurements) and the external pressure, which can be compensated if we take the deceleration into account. 
Meanwhile, a transition of the jet pressure will be expected around the structural transition, and then an overpressured jet ($p'_{jet} > p_{ext}$) can be explained. 

Nevertheless, in the structure of the NGC 6251 jet, we currently see neither a sudden decrease in the jet radius (i.e., overcollimation) nor an isolated prominent knot as a counterpart of HST-1 in the M87 jet, while there is a jagged shape in the radius profile near the SGI ($z \simeq$ 50 pc). 
It is noteworthy that studies of proper motion reported by the MOJAVE program show several knotty features moving at subluminal apparent speeds $\beta_{app} \lesssim$ 0.1 \citep{lis13}. 
Time variability may explain the current absence of such a knotty feature \citep[HST-1 is also time-varying,][]{2007ApJ...663L..65C, 2014ApJ...788..165H} and an overcollimated region at the SGI as well as the apparent jagged shape. 
Further monitoring observations with EVN and observations with higher angular resolution across the transverse direction are important for unveiling the true structure of the jet and for measuring the bulk Lorentz factor $\Gamma$. 
Polarization imaging may be helpful in probing this transition and the nature of the stationary knot. 
We also suggest that a similar analysis can be applied to M87 in order to constrain the jet pressure of the conical expansion beyond the transition.

We may also consider a smooth transition: the condition of pressure matching is met on the jet boundary ($p'_{jet} \simeq p_{ext}$) with the smooth structural transition. 
In addition to the smooth structural transition (i.e., no overcollimation), Figure \ref{fig-pressure} shows that the lower limit of the minimum-energy jet pressure seems to connect with the external pressure. 
\citet{zak08} show that an MHD jet maintains a conical shape under the condition that the external pressure has a drastically dropping profile with $b=$ 4 in the regime of constant jet speed. 
It would be difficult, however, to meet this condition. 
As we described in Figure \ref{fig-pressure}, the external pressure is nearly a constant in the conical jet regime as suggested by the estimation of the SGI and by X-ray measurements. 
In order to keep the conical (expanding) structure of the jet with the condition of pressure matching to the flat external pressure profile, in situ energy dissipation is needed to keep the jet internal pressure $p_{jet}$ constant. 
One possibility would be the conversion from the bulk kinetic energy to the jet internal energy. 
Indeed, \citet{2002MNRAS.336..328L} suggest a gradual and substantial deceleration of the jet velocity located at 1$-$3 kpc from the central engine for the FR I type jets based on the VLA polarimetric observations. 
It is interesting to measure the velocity structure of the conical jet. 
We also note that the magnetic reconnection is another possible origin for the in situ energy dissipation, and future monitoring observations of this region will detect the local time variability associated with this event. 
Therefore, further observations towards the NGC 6251 jet are needed to explore the physical origin of the structural transition in the AGN jet.

\acknowledgments
We thank M. Inoue, K. Hirotani, and all the members of Greenland Telescope (GLT)/VLBI group in ASIAA for stimulating discussions. 
We thank A. Moraghan for assistance with proofreading. 
Also, we sincerely appreciate the anonymous referee for her/his suggestions that greatly improved the manuscript. 
C.Y.T. thanks Paul T. P. Ho for warm encouragements. 
This work is partly supported by the Ministry of Science and Technology of Taiwan grants MOST 103-2112-M-001-038-MY2 (K.A.). 
The European VLBI Network is a joint facility of European, Chinese, South African, and other radio astronomy institutes funded by their national research councils. 
This research has made use of data from the MOJAVE database that is maintained by the MOJAVE team. 

\facilities{EVN, VLA, VLBA}
\software{AIPS, Difmap}



\floattable
\begin{deluxetable}{cccCCCCCC}
\tablecaption{Image Parameters
  \label{tab-1} }
\tablecolumns{9}
\tablenum{1}
\tablewidth{0pt}
\tablehead{
  \colhead{Facility} & \colhead{Code} & \colhead{Epoch} & \colhead{Frequency} & \colhead{Beam FWHM} & \colhead{RMS noise} & \colhead{Dynamic} & \colhead{Jet PA} & \colhead{${\Phi_b}$\tablenotemark{a}} \\
  \colhead{} & \colhead{} & \colhead{(yyyy-mm-dd)} & \colhead{(GHz)} & \colhead{(mas$\times$mas, deg)} & \colhead{($\mu$Jy/beam)} & \colhead{range} & \colhead{(deg)} & \colhead{(mas)}
}
\startdata
EVN-i\tablenotemark{b} & ET028 & 2013 Mar 10 & 1.658 & 3.06 $\times$ 2.96, 65$\degr$.1 & 127 & 1874 & 295.9 & 3.02 \\
EVN-ii\tablenotemark{b} & ET028 & 2013 Mar 10 & 1.658 & 4.81 $\times$ 4.01, 35$\degr$.6 & 51 & 5392 & 295.9 & 4.78 \\
EVN-iii\tablenotemark{b} & ET028 & 2013 Mar 10 & 1.658 & 16.7 $\times$ 15.8, 33$\degr$.7 & 54 & 7740 & 294.6 & 16.78 \\
VLBA & V105 & 1998 Apr 30 & 4.816 & 0.997 $\times$ 0.962, -24$\degr$.1 & 63 & 3460 & 297.9 & 0.975 \\
\sidehead{MOJAVE data}
VLBA & BT036 & 1998 Jun 02 & 15.365 & 0.450 $\times$ 0.390, -25$\degr$.8 & 204 & 1701 & 298.4 & 0.54\tablenotemark{c} \\
VLBA & BJ033B & 2000 May 30 & 15.363 & 0.468 $\times$ 0.453, 25$\degr$.6 & 454 & 844 & 298.4 & 0.54 \\
VLBA & BS089 & 2001 Feb 28 & 15.365 & 0.346 $\times$ 0.326, 34$\degr$.1 & 370 & 630 & 298.4 & 0.54 \\
VLBA & BL137R & 2007 Jan 06 & 15.365 & 0.564 $\times$ 0.468, -42$\degr$.8 & 967 & 487 & 298.4 & 0.54 \\
VLBA & BL149AA & 2007 Jun 03 & 15.365 & 0.528 $\times$ 0.500, 33$\degr$.7 & 863 & 532 & 298.4 & 0.54 \\
VLBA & BL149AF & 2007 Aug 16 & 15.365 & 0.557 $\times$ 0.481, 89$\degr$.7 & 675 & 717 & 298.4 & 0.54 \\
VLBA & BL149AK & 2008 Jul 17 & 15.365 & 0.506 $\times$ 0.494, -42$\degr$.9 & 547 & 863 & 298.4 & 0.54 \\
VLBA & BL149BF & 2008 Nov 26 & 15.357 & 0.543 $\times$ 0.483, -18$\degr$.3 & 657 & 686 & 298.4 & 0.54 \\
VLBA & BL149BM & 2009 Jun 03 & 15.357 & 0.524 $\times$ 0.496, 63$\degr$.3 & 585 & 785 & 298.4 & 0.54 \\
VLBA & BL149CL & 2010 Jul 12 & 15.357 & 0.496 $\times$ 0.482, -41$\degr$.4 & 390 & 1167 & 298.4 & 0.54 \\
VLBA & BL178AO & 2012 Sep 02 & 15.357 & 0.552 $\times$ 0.486, -6$\degr$.2 & 495 & 715 & 298.4 & 0.54 \\
VLBA & BL178BH & 2013 Aug 12 & 15.357 & 0.527 $\times$ 0.461, -25$\degr$.2 & 470 & 802 & 298.4 & 0.54 \\
\enddata
\tablenotetext{a}{$\Phi_b$: Beam FWHM perpendicular to the local jet axis.}
\tablenotetext{b}{Weighting schemes of EVN-i, -ii, and -iii images are uniform, natural, and natural with tapering, respectively.}
\tablenotetext{c}{MOJAVE images are convolved with the same circular beam of 0.54\,mas in the analysis.}
\end{deluxetable}

\floattable
\begin{deluxetable}{cccccc}
\tablecaption{Measurements of the Jet Radii and Estimated Uncertainties
  \label{tab-2} }
\tablecolumns{6}
\tablenum{2}
\tablewidth{0pt}
\tablehead{
  \colhead{$z$} & \colhead{$\Phi_0$} & \colhead{$e_{{\Phi_0}, fit}$} & \colhead{$r$} & \colhead{$e_{r, fit}$} & \colhead{$e_r$} \\
  \colhead{(mas)} & \colhead{(mas)} & \colhead{(mas)} & \colhead{(mas)} & \colhead{(mas)} & \colhead{(mas)} 
}
\startdata
\sidehead{EVN-i, 1.6 GHz, RR pol.}
16.2 & 3.73 & 0.05 & 1.09 & 0.04 & 0.61 \\
16.8 & 3.74 & 0.05 & 1.11 & 0.04 & 0.61 \\
17.4 & 3.62 & 0.05 & 1.00 & 0.04 & 0.61 \\
18.0 & 3.47 & 0.04 & 0.85 & 0.04 & 0.61 \\
18.6 & 3.34 & 0.04 & 0.71 & 0.05 & 0.61 \\
\enddata
\tablecomments{
This table is published in its entirety in the machine-readable format. 
A portion is shown here for guidance regarding its form and content. 
$z$: deprojected axial distance. 
$\Phi_0$: fitted Gaussian FWHM of the jet cross section. 
$e_{\Phi_0, fit}$: fitting uncertainties of $\Phi_0$. 
$r$: jet radius. 
$e_{r, fit}$: fitting uncertainties of $r$.
$e_{r}$: uncertainties of $r$ taking the imaging error into account, which is described in Section 3.}
\end{deluxetable}

\floattable
\begin{deluxetable}{CCCCC|c|c}
\tablecaption{Best Fit Parameters of the Power-law Models
  \label{tab-3} }
\tablecolumns{7}
\tablenum{3}
\tablewidth{0pt}
\tablehead{
  \colhead{$a$ or $a_u$} & \colhead{$A_0$} & \colhead{$a_d$} & \colhead{$z_0$}            & \colhead{$r_0$}             & \colhead{$n$} & \colhead{$\chi^2$/dof} \\
  \colhead{}         & \colhead{($r_s^{1-1/a}$)} & \colhead{}           & \colhead{(10$^5$ $r_s$)} & \colhead{(10$^3$ $r_s$)} & \colhead{}       & \colhead{}  
}
\startdata
\sidehead{Single power law,  Equation (1)}
$1.25\pm0.03$ & 0.18\pm0.05 & - & - & - & - & 0.98 \\
\hline
\sidehead{Double/Broken power law,  Equation (2)}
$6.6\pm6.0$ & - & $0.83\pm0.06$ & $1.3\pm0.7$ & $1.0\pm0.4$ & 1    & 0.58 \\
$3.0\pm0.7$ & - & $0.88\pm0.05$ & $1.7\pm0.5$ & $1.6\pm0.4$ & 2    & 0.57 \\
$2.5\pm0.4$ & - & $0.90\pm0.04$ & $1.8\pm0.5$ & $1.9\pm0.4$ & 3    & 0.57 \\
$2.3\pm0.3$ & - & $0.91\pm0.04$ & $1.9\pm0.4$ & $2.1\pm0.3$ & 4    & 0.57 \\  
$2.2\pm0.2$ & - & $0.92\pm0.04$ & $2.0\pm0.4$ & $2.2\pm0.3$ & 5    & 0.57 \\  
$2.0\pm0.1$ & - & $0.94\pm0.04$ & $2.1\pm0.3$ & $2.5\pm0.2$ & 100 & 0.57 \\ 
\enddata
\tablecomments{
$a$: power-law index of the single power-law model, 
$A_0$: normalization parameter, 
$a_u$: upstream power-law index (pre-break) of the double power-law model, 
$a_d$: downstream power-law index (post-break), 
$z_0$: deprojected axial distance of the break position, 
$r_0$: jet radius at the break position, 
$n$: controlling parameter for the sharpness of the break,
and $\chi^2$/dof: reduced chi-square of the fitting.
}
\end{deluxetable}



\end{document}